\documentclass[pra,aps,showpacs]{revtex4}
\usepackage{amsthm,amssymb,amsmath,amsfonts}
\usepackage{epsfig}
\newcommand{\beao}{\begin{eqnarray*}}
\newcommand{\eeao}{\end{eqnarray*}}
\newcommand{\be}{\begin{equation}}
\newcommand{\ee}{\end{equation}}
\newcommand{\bea}{\begin{eqnarray}}
\newcommand{\eea}{\end{eqnarray}}
\newcommand{\nn}{\nonumber}

\newcommand{\Ref}[1]{(\ref{#1})}
\newcommand{\F}{{\cal F}}
\newcommand{\Tr}{\rm Tr}
\begin{document}
\title{On the Casimir repulsion in sphere-plate geometry}
\date{\today}
\author{Irina G. Pirozhenko\footnote{pirozhen@theor.jinr.ru}}
\affiliation{
Bogoliubov Laboratory of Theoretical Physics, Joint Institute for
Nuclear Research, Dubna 141 980,
Russia}
\altaffiliation{
Dubna International University, Dubna, Russia}
\author{Michael Bordag\footnote{Michael@Bordag.com}}
\affiliation{Universit{\"a}t Leipzig, Institute for Theoretical Physics, Postfach 100920, 04009 Leipzig, Germany}
\begin{abstract}
The electromagnetic vacuum energy is considered in the presence of a perfectly conducting plane and a ball with dielectric permittivity $\varepsilon$ and magnetic permeability $\mu$, $\mu\ne1$. The attention is focused on the Casimir repulsion in this system caused by magnetic permeability of the sphere. In the case of perfectly permeable sphere, $\mu=\infty$, the vacuum energy is estimated numerically.
The short and long distance asymptotes corresponding to the repulsive force and respective low temperature corrections and high temperature limits are found for a wide range of $\mu$. The constraints on the Casimir repulsion in this system are discussed.
\end{abstract}
 
\maketitle
\section{Introduction}
The electromagnetic vacuum energy in the sphere-plate geometry has been intensively studied since 2007~\cite{Emig:PhysRevLett.99.170403,Emig:2008,PhysRevA.78.012115,PhysRevLett.102.230404}. 
The analytic corrections to the proximity force approximation were first found in~\cite{Bordag:PhysRevD.81.065011}, the long distance (small sphere) asymptote was derived in~\cite{Emig:PhysRevLett.99.170403}. Then the attention turned to geometry-temperature interplay in the system~\cite{Gies:arXiv:1003.3420,Gies:Int.J.Mod.Phys.A,Canaguier:PhysRevLett.104.040403,Bordag:PhysRevD.81.085023}.  In the above mentioned papers the plate was perfectly conducting while on the sphere the electromagnetic field obeyed either boundary conditions of a perfect conductor or a dielectric with constant or frequency dependent dielectric permittivity and magnetic permeability.  In~\cite{Milton:2009gk} semitransparent boundary conditions were imposed.

For parallel plates the Casimir repulsion is expected if one plate is mainly dielectric and the other is mainly magnetic.
The strongest repulsion  is achieved for perfectly permeable plate, $\mu\to\infty$, facing perfectly conducting one, $\varepsilon\to\infty$.  The magnitude of the repulsive force can not exceed $7/8$ of the Casimir force between perfectly conducting plates. 
This constraint for the repulsion was obtained by Boyer~\cite{Boyer:1974} in 1974.
At finite temperature the repulsion in plate-plate geometry was studied in~\cite{PhysRevD.60.105022,0295-5075-72-6-929}.  In the high temperature limit the repulsive force can not exceed $-3/4 f_T$, where $f_T$ is the high temperature limit of the Casimir force between two perfectly conducting plates.

To our knowledge the problem of Casimir repulsion between non-planar objects was touched upon in the scattering approach~\cite{PhysRevD.80.085021} though with more attention payed to repulsion in cylinder-plate geometry.  
At the same time the metallic bodies with more sophisticated shape were studied, and for certain cases Casimir repulsion due to the geometry was predicted in~\cite{PhysRevLett.105.090403}. The repulsion  for fluid-separated sphere and plate  with $\varepsilon_{sph}(i\xi)<\varepsilon_{fluid}(i\xi)<\varepsilon_{plate}(i\xi)$ was considered in~\cite{PhysRevLett.104.160402,PhysRevA.83.052503}.

In the present paper we return to the repulsion in sphere-plate geometry and our aim is to find the constraints on the repulsion caused by magnetic permeability of the sphere. We do not develop new methods of calculation here, the approaches of the papers~\cite{Emig:PhysRevLett.99.170403,Emig:2008,PhysRevA.78.012115,PhysRevLett.102.230404,Canaguier:PhysRevLett.104.040403,Bordag:PhysRevD.81.085023,Bordag:2010qq} are used . The short and long distance asymptotes corresponding to the  repulsive force and respective low and high temperature corrections are found analytically. The energy at medium distances and temperatures is evaluated numerically.

The outline of the paper is the following. Section II comprises the basic formulas  used throughout the paper and presents the numerical constraints for the Casimir repulsion  at zero temperature. In Section II we present long and short distance asymptotes of the vacuum energy.  High and low temperatures  corrections are derived in Section III, where also the influence of the temperature on the repulsion is discussed. 

\noindent Throughout the paper we use units with  $\hbar=c=k_B=1$.

\section{Basic formulae and numerical constraints for the Casimir energy }
We begin this section with the main formulas for the computation of the electromagnetic vacuum energy in sphere-plane geometry.
The functional integral quantization of the electromagnetic field  with account for boundary conditions yields the following expression for the the vacuum energy~\cite{BKMM,Emig:2008},
\begin{equation}
\label{E1}
E_0=\frac{1}{2}\int_{-\infty}^\infty\frac{d \xi}{2\pi}
\Tr \ln \left(1-\mathbf{M}(\xi)\right).
\end{equation}
This is the separation dependent part of the vacuum energy. It is finite, as the infinite vacuum energy of the electromagnetic field  with  bodies placed at infinite separation has been subtracted. 
Here $\mathbf{M}$ is a matrix in orbital momentum index, $M_{l,l'}$, $1<l,l'<\infty$; moreover, $\mathbf{M}$ is a diagonal matrix in {\it magnetic quantum number} $m$, and at the same time $\mathbf{M}$ is a $(2\times2)$ matrix in electromagnetic polarizations,
\bea\label{1.NED} \mathbb{M}_{l,l'}(\xi)
&=&
\sqrt{\frac{\pi}{4 \xi L }}\sum_{l''=|l-l'|}^{l+l'}K_{l''+1/2}(2 \xi L
) H_{ll'}^{l''}\,
\\&&\nn
\times\left(\begin{array}{cc}\Lambda_{l,l'}^{l''}&\tilde{\Lambda}_{l,l'}
\\ \tilde{\Lambda}_{l,l'}&\Lambda_{l,l'}^{l''}\end{array}\right)
\left(\begin{array}{cc}d^{\rm TE}_l(\xi R)&0
\\ 0&-d^{\rm TM}_l(\xi R)\end{array}\right)
 \,.
\eea
The indices $TE$ and $TM$ refer to transverse electric and transverse magnetic polarizations. The functions  $d^{\rm TE}_l(\xi R)$ and $d^{\rm TM}_l(\xi R)$ originate from the T-matrix for the scattering on the sphere.
For a sphere with dielectric permittivity $\varepsilon$ and magnetic permeability $\mu$ they are given by
\be \label{3.E}
d_{l}^{TE}(z)=\frac{2}{\pi} \frac{\sqrt{\varepsilon}s_{l}(z)s_{l}^{'}(n z)-\sqrt{\mu}s_{l}^{'}(z)s_l(n z)}{\sqrt{\varepsilon}e_{l}(z)s_{l}^{'}(n z)-\sqrt{\mu}e_{l}^{'}(z)s_l(n z)},
\ee
where $n=\sqrt{\varepsilon \mu}$.
Here $s_l(z)$ and $e_l(z)$ are modified  Riccati-Bessel functions:
\begin{equation}
s_l(z)=\sqrt{\frac{\pi z}{2}} I_{l+1/2}(z), \quad e_l(z)=\sqrt{\frac{2 z}{\pi}} K_{l+1/2}(z).
\end{equation}

To obtain $d_{l}^{TM}(\xi)$ one has to exchange  $\mu$ and $\varepsilon$ in~(\ref{3.E}). 

For perfect conductor boundary conditions on the sphere, $\varepsilon\to\infty$,  $d^{\rm TE}$ and  $d^{\rm TM}$ simplify to 
\bea\label{1.dTE} d^{\rm TE}_l(z)&=& \frac{I_{l+1/2}(z)}{K_{l+1/2}(z)}\\
\label{1.dTM} d^{\rm TM}_l(\xi R)&=& \frac{\left(I_{l+1/2}(z)\sqrt{z}\right)'} {\left(K_{l+1/2}(z)\sqrt{z}\right)'}\,.
\eea
For the boundary conditions on perfect magnetic body,  $\mu\to\infty$, on has to replace $d^{TE}\leftrightarrow d^{TM}$ in (\ref{1.NED}).
The derivation of this formulas can be  found, for example, in \cite{Emig:2008,BKMM}. In the present paper we use the notations of~\cite{Bordag:PhysRevD.81.085023}.

The numerical factors $H_{ll'}^{l''}$ in \Ref{1.NED} are given by
\bea\label{1.H} 
H_{ll'}^{l''}&=&  \sqrt{(2l+1)(2l'+1)}(2l''+1)\times
\left(\begin{array}{ccc}l&l'&l''\\0&0&0\end{array}\right)
\left(\begin{array}{ccc}l&l'&l''\\m&-m&0\end{array}\right), \eea
where the parentheses denote the $3j$-symbols. 
The so-called translation formulas for a vector field~\cite{BKMM} produce  the multipliers
\bea\label{2.LA}
\Lambda_{ll'}^{l''}=\frac{\frac12\left[l''(l''+1)-l(l+1)-l'(l'+1)\right]}
{\sqrt{l(l+1)l'(l'+1)}}, \quad
\tilde{\Lambda}_{ll'}=\frac{2m\xi L}{\sqrt{l(l+1)l'(l'+1)}}.
\eea

After  turning to dimensionless integration  variable, $x\to \xi L$, in   (\ref{E1})  one  can obtain $E_0$ for any finite $0<\rho<1$, $\rho=R/L$, numerically. At large $x$ the integrand in (\ref{E1}) behaves as 
$\Tr \ln (1-\mathbf{M}) \sim \pm \exp[2(\rho-1)x]/x$, therefore at any nonzero separation between the sphere and the plate, 
$0<\epsilon<1$, the integral converges at the upper limit. In the limit $\rho\to1$ the PFA becomes valid.  For a given separation $\rho$ the matrix $\mathbf{M}(0)$ may be truncated at $l=\lambda(\rho)$ such that $|\Tr\ln(1-\mathbf{M}(0))_{\lambda}-\Tr\ln (1-\mathbf{M}(0))_{\lambda+1}|<\delta$. One can as well calculate the $E_0(\rho, \lambda)$ for a sequence of $\lambda$ and then extrapolate the the result to $\lambda\to\infty$~\cite{Emig:2008}.  

\begin{figure}[b,t]
\epsfig{file=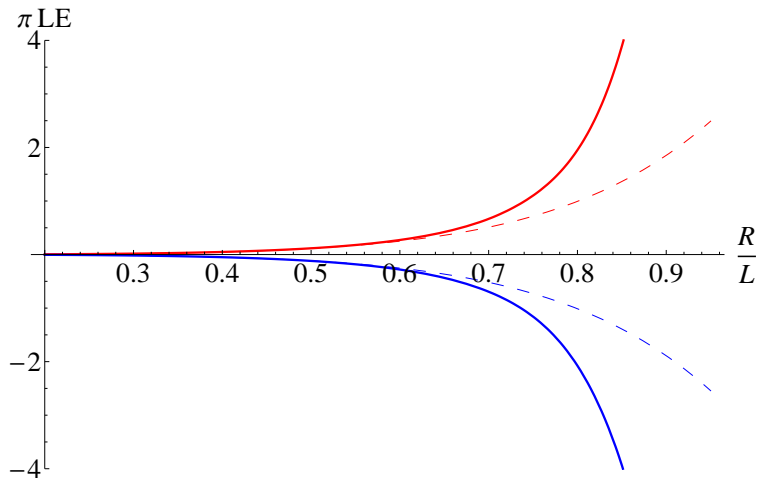,width=10cm}
\caption{The vacuum energy giving contribution to  the force between the sphere and the plane takes values in the  region between the red and blue solid lines corresponding respectively to the strongest possible repulsion and attraction. The dashed lines give the large separation asymptotes of the energy. The ratio of the strongest possible repulsion to the strongest possible attraction depends on the separation between the sphere and the plate and changes from -7/8 at short distances to -1 at long distances. At medium distances the energy was estimate numerically. For example, when $R/L=0.5$, $E^{rep}/E^{attr}\simeq-0.98$; for $R/L=0.8$,  $E^{rep}/E^{attr}\simeq-0.94$.}
\label{Fig1}
\end{figure}

The  numerical constraints  for the vacuum energy   in the presence of the sphere facing the perfectly conducting plate at $T=0$ are presented in Fig.~\ref{Fig1}.  The boundary conditions on the sphere (or its material) define the value and sign of the vacuum energy.  
It may vary from the vacuum energy in the presence of perfectly conducting sphere  which yields the strongest attraction, to the vacuum energy in the presence of perfectly permeable sphere, providing the strongest repulsion. 

The transition to finite temperature is achieved through the Matsubara formalism. The free energy is expressed as a sum over Matsubara frequencies 
 $\xi_n=2\pi Tn$, where $T$ -  is the temperature of the system 
\be
\label{Matsub} \F=\frac{T}{2}\sum_{n=-\infty}^{\infty} \Tr \ln
\left(1-\mathbf{M}(\xi_n)\right).
\ee
The low, medium and high temperatures for the system  are defined with respect to the radius of the sphere $R$ and the distance between the sphere and the plate $d$. At low temperature the inequality  $T<<1/R, 1/d$ holds,  
high temperature is defined by   $T>> 1/R, 1/T $~\cite{Bordag:PhysRevD.81.085023}.

\section{Large and short separation asymptotes}
\subsection{Large separations}
When the  sphere and the  plane are far apart, the ratio  $\rho=R/L$ tends to zero. Following~\cite{Emig:2008} we expand to logarithm in~(\ref{E1}) up to the third term and truncate the matrix  $\mathbf{M}$ at $l=4$. Then  the vacuum energy at large separations is a power series with respect to $\rho$,
\begin{equation}
E_0=\frac{1}{\pi L}\sum_{j=4}^{\infty}c_j \rho^{j-1}.
\end{equation}    
The coefficients $c_j$, $j=4..11$ for perfectly conducting sphere were found in~\cite{Emig:2008}. 
  Here we present the coefficients for perfectly permeable sphere with $\mu\to\infty$,
\begin{eqnarray}
c_4=\frac{9}{16},\quad c_5=0, \quad c_6=\frac{25}{32}, \quad
c_7=\frac{2737}{4096}, \quad
c_8=\frac{12551}{9600}, \nonumber \\
c_9=-\frac{1298187}{163840}, \quad
c_{10}=\frac{31982323007}{722534400}, \quad
c_{11}=-\frac{39548025347}{412876800}.
\end{eqnarray}

The coefficients $c_4,\; c_5,\; c_6$ coincide in magnitude with those derived for perfectly conducting sphere, however they differ in sign. 
At large separations the vacuum energy in the presence of perfectly conducting plane and perfectly permeable sphere is positive corresponding to repulsion. 

For a sphere with constant dielectric permittivity and magnetic permeability the coefficients read
 
\begin{eqnarray}
&& c_4=-\frac{9 (\varepsilon-\mu)}{8 (2+\varepsilon) (2+\mu)}, \quad c_5=0, \nonumber\\
&&c_6=-\frac{3 (\varepsilon-\mu) \left(380+320 (\varepsilon+\mu)+50(\varepsilon^2+\mu^2)+217 \varepsilon \mu+8 \varepsilon \mu (\varepsilon+ \mu)-11 \varepsilon^2 \mu^2+3 \varepsilon^2 \mu^2(\varepsilon+\mu)+2 \varepsilon^3 \mu^3\right)}{8 (2+\varepsilon)^2 (3+2 \varepsilon) (2+\mu)^2 (3+2 \mu)}, \nonumber\\
&&c_7=\frac{128-11584 \varepsilon+3023 \varepsilon^2+11456 \mu-382 \varepsilon \mu+5792 \varepsilon^2 \mu-2737 \mu^2-5728 \varepsilon \mu^2+32 \varepsilon^2 \mu^2}{1024 (2+\varepsilon)^2 (2+\mu)^2}. \nonumber
\end{eqnarray} 
   
For a sphere with dielectric permittivity and magnetic permeability defined by the plasma model, $\varepsilon(\xi)=1+\omega_p^2/\xi^2$, $\mu(\xi)=1+\omega_m^2/\xi^2$, the coefficients depend on $\omega_p$ and $\omega_m$,
\begin{eqnarray}
&&c_4=0, \quad  c_5=-\frac{9  (\omega_p^2-\omega_m^2)}{16 {\omega_m} {\omega_p}}, \quad  \
c_6=\frac{15 (\omega_p^4-\omega_m^4)}{16 \omega_m^2 \omega_p^2}, \nonumber \\
&&c_7=- \frac{1}{32}-\frac{135}{64} \frac{(\omega_p^6-\omega_m^6)}{\omega_m^3 \omega_p^3}
- \frac{415}{256} \frac{(\omega_m^2 - \omega_p^2) }{256 \omega_m \omega_p}
 -\frac{135}{64 R^4} \frac{(\omega_p^2 -\omega_m^2)}{\omega_m^3 \omega_p^3}
 + \frac{125}{128 R^2}\frac{(\omega_p^4-\omega_m^4)}{\omega_m^3 \omega_p^3}.  
\end{eqnarray}

\subsection{Small separations}
At small separations  the leading contribution to the energy is determined by the PFA. The corrections to PFA may be found either by derivative expansion of the vacuum energy~\cite{Fosco:2011, Fosco:2012, BimonteEmig:2011,BimonteEmig:2012} or by summing the asymptotic scattering series at short distances~\cite{Bordag:PhysRevD.73.125018,1751-8121-41-16-164002,PhysRevD.84.125037}. At short distances the parameter $\rho\to 1$ is not small therefore the corrections to PFA are sought in terms of expansion in powers of $d/R$.

It was shown in~\cite{PhysRevD.84.125037}  that the first two leading terms of the electromagnetic Casimir energy can be written as 
\begin{eqnarray}
E^{cond}=\left\{\begin{array}{c} \text{first two leading} \\ \text{terms of } E^{DD}\end{array} \right\}+
\left\{\begin{array}{c}  \text{first two leading} \\ \text{terms of } E^{NR}|_{u=1/2}\end{array} \right\}+\Delta E,
\label{PFA1}
\end{eqnarray}
with $E^{XY}$ being  the vacuum energy of a massless scalar field, where $X$ denotes the boundary conditions on the plane, and $Y$ stands for the boundary conditions on the sphere. The detailed derivation for the
Dirichlet, $X=D$ (or Neumann, $X=N$), boundary conditions on the plane and Dirichlet, $Y=D$ (or Robin, $Y=R$), boundary conditions on the sphere was performed in~\cite{Bordag:PhysRevD.81.065011,PhysRevD.84.125037}. The results for the scalar field with various boundary conditions are the following, 
\begin{eqnarray}
E^{DD}&=&-\frac{\pi^3 R}{1440 d^2}\left(1+\frac{1}{3}\frac{d}{R}+\dots\right), \nonumber \\
E^{ND}&=&\frac{7\pi^3 R}{11520 d^2}\left(1+\frac{1}{3}\frac{d}{R}+\dots\right), \nonumber\\
E^{NR}&=&-\frac{\pi^3 R}{1440 d^2}\left(1+\left[\frac{1}{3}+\frac{10(6u-1)}{\pi^2}\right]\frac{d}{R}+\dots\right), \nonumber\\
E^{DR}&=&\frac{7\pi^3 R}{11520 d^2}\left(1+\left[\frac{1}{3}+\frac{40(6u-1)}{7\pi^2}\right]\frac{d}{R}+\dots\right).
\label{PFA_scal}
\end{eqnarray}
The Robin parameter $u$ equals to $-1/2$ for Neumann boundary conditions.

The $\Delta E$ term is defined by the factors $\Lambda$ and $\tilde{\Lambda}$~(\ref{2.LA}). In other words, it is governed by the geometry of the system, but does not depend on the type of boundary conditions. It was found in~\cite{PhysRevD.84.125037},
\begin{equation}
\Delta E = \frac{R}{4\pi d^2}\frac{\pi^2}{6}\frac{d}{R}.
\label{deltaE}
\end{equation}

Combining~(\ref{PFA_scal}) and (\ref{deltaE}) according to (\ref{PFA1}) one obtains for a perfectly conducting sphere in front of a perfectly conducting plane
\begin{equation}
E^{cond}=-\frac{\pi^3 R}{720 d^2}\left(1+\left[\frac{1}{3}-\frac{20}{\pi^2} \right]\frac{d}{R}+\dots\right)\simeq E_{PFA}\left[1-1.69 \frac{d}{R}\right].
\end{equation} 

For perfect magnetic  boundary conditions  on the  sphere  one just has to exchange $d^{TE}$ and $d^{TM}$ in the matrix $M$. Keeping this in mind  we construct the leading terms beyond PFA for this system in analogy with~(\ref{PFA1}) using the scalar field results (\ref{PFA_scal}) and correction (\ref{deltaE}), 
\begin{eqnarray}
E^{magn}=\left\{\begin{array}{c} \text{first two leading} \\ \text{terms of } E^{DR}|_{u=1/2}\end{array} \right\}+
\left\{\begin{array}{c}  \text{first two leading} \\ \text{terms of } E^{ND}\end{array} \right\}+\Delta E.
\label{PFA2}
\end{eqnarray}
Substituting (\ref{PFA_scal}) and (\ref{deltaE}) into (\ref{PFA2}) one derives the short distance asymptote of the electromagnetic vacuum energy in the presence of  perfectly permeable ball and  perfectly conducting plane
\begin{equation}
E^{magn}=\frac{7\pi^3 R}{5760 d^2}\left(1+\left[\frac{1}{3}+\frac{40}{\pi^2} \right]\frac{d}{R}\right)\simeq -\frac{7}{8} E_{PFA}\left[1+4.38 \frac{d}{R}\right].
\end{equation} 

\section{High temperature limit and low temperature corrections}

\subsection{High temperature}
The high temperature limit for the Casimir force in plate-plate geometry was discussed in a number of papers and reviewed in~\cite{BKMM}. At high temperature, the force per unit area satisfies  the condition: $f_T\leq f<-3/4\; f_T$,  $f_T\equiv-\zeta(3) T/(8\pi d^2)$. In sphere-plate geometry the high-temperature (classical) limit with account for finite conductivity was studied, for example, in~\cite{Durand:2012}. 

The high-temperature behavior of the free energy follows from the Matsubara sum \Ref{Matsub}. The leading classical term at $T\to\infty$ is defined by the lowest Matsubara frequency. Separating this term (with $n=0$), we denote the remaining ones by $F_1$
\be\label{1.HT}
\F=T \, F_0(\rho)+F_1(Td,TR).
\ee
The leading contribution is proportional to $T$ and $F_0$ depends only on  $\rho$. The function  
 $F_1$ depends on two dimensionless combinations. 

\begin{figure}[b,t]
\epsfig{file=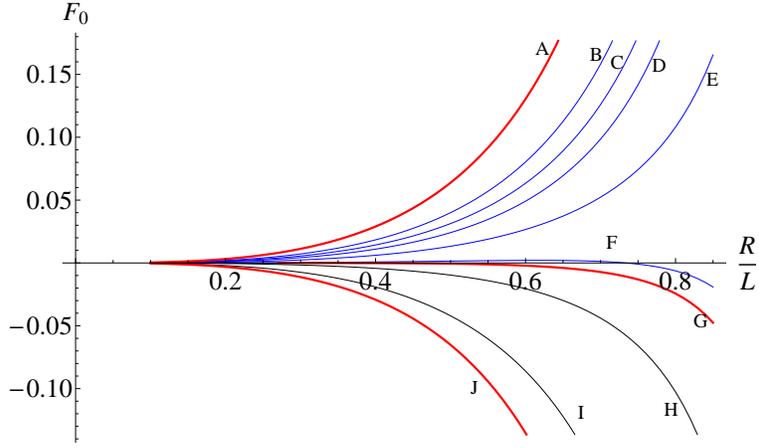,width=10cm}
\caption{High temperature limit. $F_0$ as a function of $R/L$. The plane is perfectly conducting. The curves correspond to different values of the parameters related to the ball:  (A) perfectly permeable ball; (B) $\varepsilon=1$, $\mu=100$; (C) $\varepsilon=1$, $\mu=10$; 
(D) $\varepsilon=1$, $\mu=5$; (E) $\varepsilon=2$, $\mu=5$; (F) $\varepsilon=8$, $\mu=10$; (G) $\varepsilon=\mu=10$; (H) $\varepsilon=100$, $\mu=10$;
(I) $\varepsilon=1000$,$\mu=1$; (J) perfectly conducting ball.}
\label{Fig2}
\end{figure}

The function $F_0$ is defined by 
\be
\label{2.HT}
F_0(\rho)=\frac{1}{2}\Tr \ln (1-\mathbf{M}(0)).
\ee
For a perfectly conducting ball, 
\be\label{3.HT_cond}
M_{l,l'}^{\rm TM}(0)=\frac{l+1}{l} \tilde{M}_{l,l'}(0), \quad M_{l,l'}^{\rm TE}(0)=\tilde{M}_{l,l'}(0).
\ee
At large separations  one arrives at  $F_0=-3/8 \rho^3 + \mathcal{O}(\rho^5)$. See curve (J), Fig.~\ref{Fig2}, for numerical results with
$0.1<\rho<0.8$.

For a perfectly magnetic ball in front of a perfectly conducting plate one has
\be\label{3.HT_magn}
M_{l,l'}^{\rm TE,pm}(0)=-\frac{l+1}{l} \tilde{M}_{l,l'}(0), \quad M_{l,l'}^{\rm TM,pm}(0)=-\tilde{M}_{l,l'}(0), \\
\ee
where
\be
\tilde{M}_{l,l'}(0)=(-1)^{l'} \frac{\sqrt{\pi}}{2}H^{l+l'}_{ll'}\Lambda^{l+l'}_{ll'}\frac{(R/2)^{2l+1}}{L^{l+l'+1}}\frac{\Gamma(-l+\frac{1}{2})}{\Gamma(-l-l'+\frac{1}{2})\Gamma(l+\frac{3}{2})}.
\label{4.HT}
\ee
At large separations  $F_0=3/8 \rho^3 + \mathcal{O}(\rho^5)$ holds. The numerical results with
$0.1<\rho<0.8$ are comprised in curve (A), Fig.~\ref{Fig2}.

For the material ball with $d_l(\xi)$ defined by (\ref{3.E}), three cases are of interest,
\bea
({\rm \bf a})&& \varepsilon(z)=\varepsilon_0, \; \mu(z)=\mu_0,  \quad  nz|_{z\to0}\sim\sqrt{\varepsilon_0\mu_0}\,z, \quad \left. s_{l}^{'}(n z)/s_{l}(n z)\right|_{z\to0}\sim\frac{(l+1)}{\sqrt{\varepsilon_0\mu_0}\,z}, \nonumber\\
({\rm \bf b})&&\varepsilon(z)=1+\frac{\omega_p^2}{z^2}, \; \mu(z)=\mu_0,  \quad nz\sim \omega_p \sqrt{\mu_0}, \nonumber\\
&or& \; \varepsilon(z)=\varepsilon_0, \; \mu(z)=1+\frac{\omega_m^2}{z^2}, \, \quad nz\sim \omega_m \sqrt{\varepsilon_0},
 \quad \left. s_{l}^{'}(n z)/s_{l}(n z)\right|_{z\to0}\to const,  \nonumber \\
({\rm \bf c})&&\varepsilon(z)=1+\frac{\omega_p^2}{z^2}, \, \mu(z)=1+\frac{\omega_m^2}{z^2}, \quad nz\sim \omega_p\omega_m /z,
\quad \left. s_{l}^{'}(n z)/s_{l}(n z)\right|_{z\to0}\to 1.  
\label{5.HT}
\eea 

Let us consider these cases in detail.

({\rm \bf a}) For constant dielectric permittivity and magnetic permeability, 
\be\label{6.HT}
M_{l,l'}^{\rm TE}(0)=\tilde{M}_{l,l'}(0)\frac{(l+1)(1-\mu)}{(l+1)+\mu l}, \quad 
M_{l,l'}^{\rm TM}(0)=-\tilde{M}_{l,l'}(0)\frac{(l+1)(1-\varepsilon)}{(l+1)+\varepsilon l}. 
\ee
Substituting these expressions into (\ref{2.HT}) one obtains $F_0$ for any finite $\epsilon=R/L$ numerically.    Fig. \ref{Fig2} gives  $F_0$ as a function of $R/L$. The curves (A)-(D) refer to the ball with constant dielectric permittivity and magnetic permeability, and $\mu>\varepsilon$. In these systems the leading term of the free energy is positive and we expect repulsion between the sphere and the plate.  The curve (F) with $\varepsilon/\mu=0.8$ crosses the abscissa at $R/L\sim0.74$. For the curves (G)-(J) $\mu\le\varepsilon$. For them the leading term of the free energy~(\ref{1.HT}) is negative indicating attraction between the sphere and the plate. 

%

At large distances the '$\Tr \ln$' in (\ref{2.HT}) may be expanded in powers of $\rho$. The leading term is defined by the lowest orbital numbers
\begin{eqnarray}
\label{7.HT}
\Tr \ln(1-\mathbf{M}(0))&\approx & -\Tr \,\mathbf{M}(0) \approx - \sum\limits_{TE,TM}^{}\left[M^{m=0}_{1,1}(0)+ 2 M^{m=1}_{1,1}(0)\right] =   -\frac{1}{2}\left(\frac{1-\mu}{2+\mu}-\frac{1-\varepsilon}{2+\varepsilon}\right) \rho^3.
\end{eqnarray}
The leading order of  the free energy $\F$ is then obtained by multiplying \Ref{7.HT} with  $T/2$,
\be
\label{8.HT}
\F=-\frac{T}{4}  \left(\frac{1-\mu}{2+\mu}-\frac{1-\varepsilon}{2+\varepsilon}\right)\rho^3+F_1(Td,TR).
\ee

({\rm \bf b}) For the dielectric permittivity defined by plasma model and constant magnetic permeability,
\be\label{9.HT}
M_{l,l'}^{\rm TE}(0)=\tilde{M}_{l,l'}(0)\frac{\Omega_p \, s_{l}'(\Omega_p \sqrt{\mu}_0)-\sqrt{\mu}_0\, s_{l}(\Omega_p \sqrt
{\mu}_0) \,(l+1)}{\Omega_p \; s_{l}'(\Omega_p\sqrt{\mu}_0)+\sqrt{\mu}_0 \; s_{l}(\Omega_p\sqrt{\mu}_0) \, l}, \quad 
M_{l,l'}^{\rm TM}(0)=\tilde{M}_{l,l'}(0)\frac{l+1}{l}. 
\ee
When $\Omega_p\to\infty$,  we reproduce the result for perfectly conducting ball~(\ref{3.HT_cond}).
The corresponding $F_0$ is plotted  in Fig. \ref{Fig2} by the curve (J).
At short distances the free energy may be estimated by high temperature PFA  developed in~\cite{Bordag:PhysRevD.81.085023}, 
$\F=-\zeta(3)RT/(4d)$. At large distances the formulas~(\ref{9.HT}) yield the following analytic expression,
\begin{equation}
F_0(\rho) \approx -\frac{3}{8}\frac{-\sqrt{\mu_0} \Omega_p +\left(1+\mu_0 \Omega_p^2\right) \text{th}\left(\sqrt{\mu_0} \Omega_p\right)}{(\mu_0-1) \sqrt{\mu_0} \Omega_p +\left((1 +\mu_0\Omega_p^2) -\mu_0 \right) \text{th}\left(\sqrt{\mu_0} \Omega_p\right)}\rho^3+\mathcal{O}\left(\rho^5\right).
\label{10.HT}
\end{equation}
Here $\Omega_p=\omega_p R=2\pi R/\lambda_p$ and $\Omega_m=\omega_m R=2\pi R/\lambda_m$.

Similarly, for constant dielectric permittivity and magnetic permeability defined by the plasma model,
\be\label{11.HT}
M_{l,l'}^{\rm TE}(0)=-\tilde{M}_{l,l'}(0)\frac{l+1}{l}, \quad
M_{l,l'}^{\rm TM}(0)=-\tilde{M}_{l,l'}(0)\frac{\Omega_m s_{l}'(\Omega_m \sqrt{\varepsilon}_0)-\sqrt{\varepsilon}_0 s_{l}(\Omega_m \sqrt{\varepsilon}_0)(l+1)}{\Omega_m s_{l}'(\Omega_m\sqrt{\varepsilon}_0)+\sqrt{\varepsilon}_0s_{l}(\Omega_m\sqrt{\varepsilon}_0)l}. 
\ee
When $\Omega_m\to\infty$  we reproduce the result for perfectly permeable ball~(\ref{3.HT_magn}).
The corresponding $F_0$ is plotted  in Fig. \ref{Fig2} by the curve {\bf A}.
At short distances the PFA yields $\F=3 \zeta(3)RT/(16 d)$. At large distances  one can derive 
\begin{equation}
F_0(\rho)\approx \frac{3}{8}\frac{-\sqrt{\varepsilon_0} \Omega_m +\left(1+\varepsilon_0 \Omega_m^2\right) \text{th}\left(\sqrt{\varepsilon_0} \Omega_m\right)}{(\varepsilon_0-1) \sqrt{\varepsilon_0} \Omega_m +\left((1 +\varepsilon_0\Omega_m^2) -\varepsilon_0 \right) \text{th}\left(\sqrt{\varepsilon_0} \Omega_p\right)}\rho^3+\mathcal{O}\left(\rho^5\right).
\label{12.HT}
\end{equation}
Again, we observe that at high temperature the ratio between the strongest possible repulsion and the strongest possible attraction 
depends on the distance between the sphere and the plate, varying from -3/4 at short separations to -1 at long separations.

\begin{figure}[b,t]
\epsfig{file=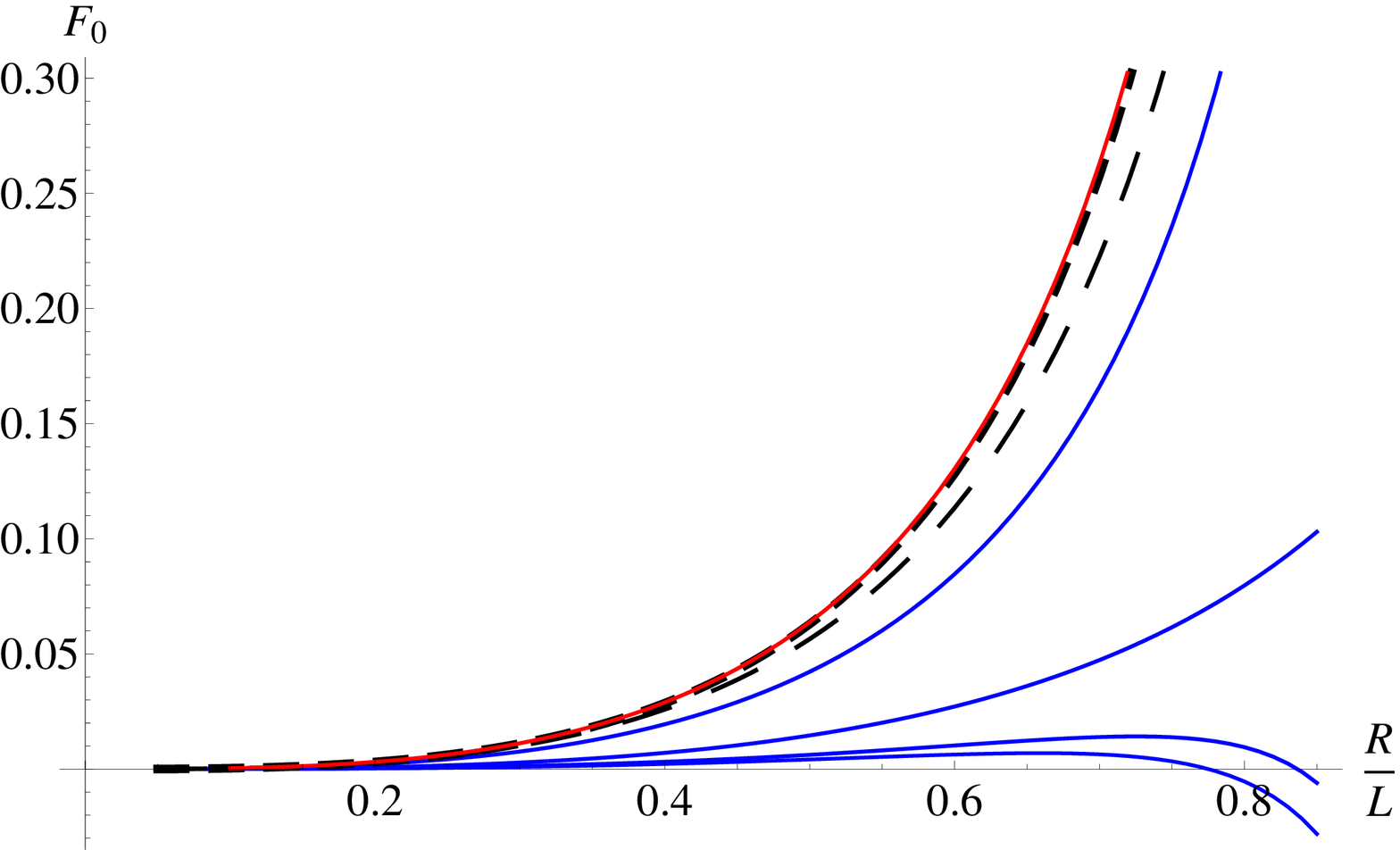,width=8cm}
\epsfig{file=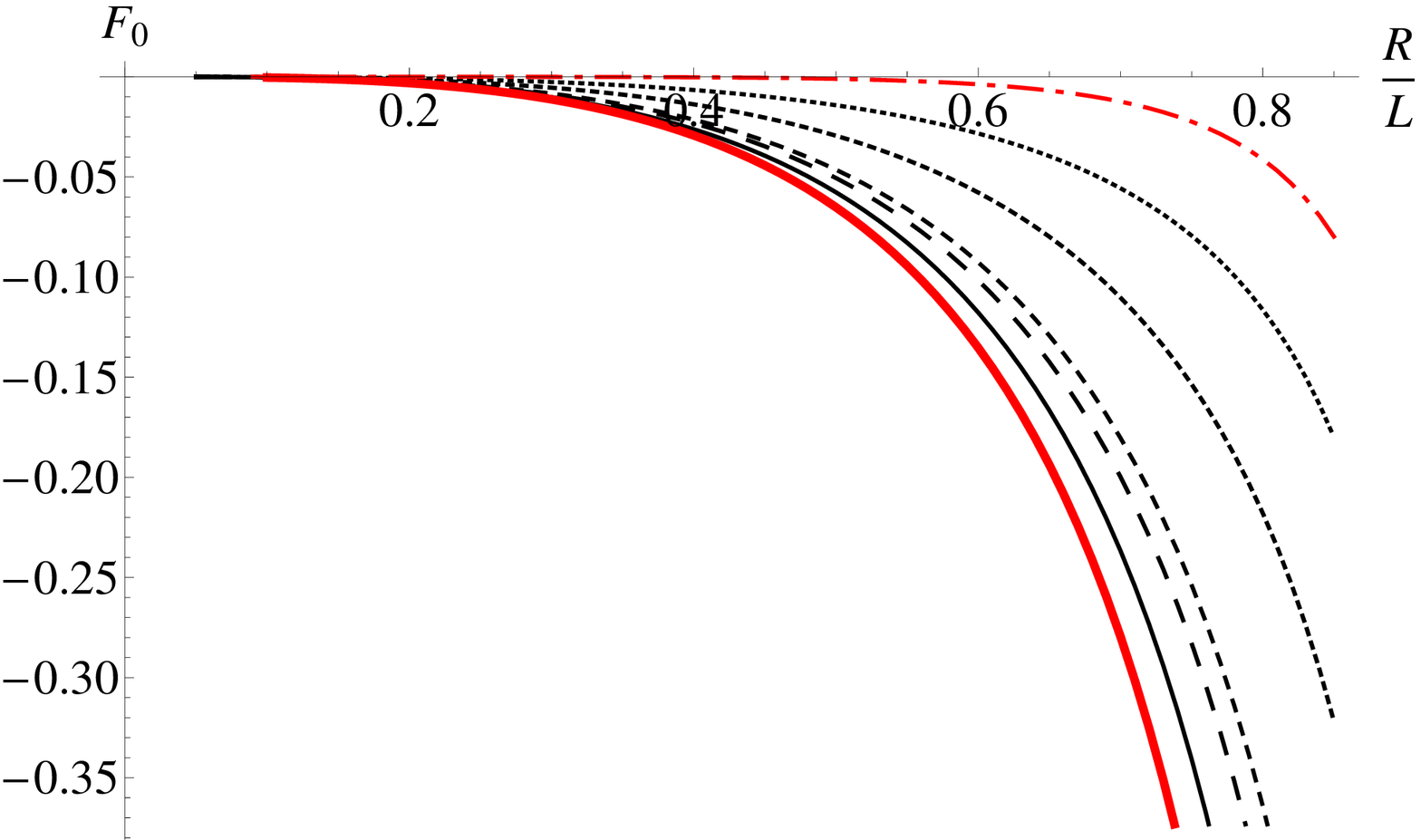,width=8cm}
\caption{$F_0$ as a function of the radius $R/L$ for varying $\omega_p$ and $\mu_0$ (or varying $\omega_m$ and $\varepsilon_0$).
The {\bf left panel} gives the plots for a ball with constant dielectric permittivity $\varepsilon$ and magnetic permeability defined by plasma model. The red  curve corresponds to perfectly permeable ball. The  blue curves are obtained for $\Omega_m=1$ and $\varepsilon=1,10,50,100$. The larger the constant dielectric permittivity  is, the lower goes the curve. When $\varepsilon=50$ and $100$, the $F_0$ crosses the abscissa, therefore the energy changes the sign.  The black dashed curves refer to the ball with $\Omega_m=10,100$ and $\varepsilon=1$. When $\Omega_p=100$ the plot approaches the one for perfectly permeable ball. The {\bf right panel} presents the plots for a ball with constant magnetic permeability $\mu$ and dielectric permittivity defined by plasma model. The red thick curve corresponds to perfectly conducting ball. The red thick dash-dotted curve is a plot for the ball with both dielectric permittivity and magnetic permeability defied by plasma model.  For the other curves $\Omega_p=10$,
$\mu=1,5,10,100,1000$. With increasing $\mu$ the absolute value of $F_0$ decreases.}
\label{Fig3}
\end{figure}

Changing the dielectric permittivity or magnetic permeability one can change the sign of the force.  Fig. \ref{Fig3} presents the results of numerical calculation for case (b). We fix the radius of the ball and plot $\Tr\ln(1-M(0))$ as a function of  $R/L$ for varying $\omega_p$ and $\mu_0$ (left panel) and  varying $\omega_m$ and $\varepsilon_0$ (right panel). 

({\rm \bf c}) When both, $\varepsilon$ and $\mu$, are defined by plasma model, $M_{l,l'}^{\rm TM}(0)$ and $M_{l,l'}^{\rm TE}(0)$ do not depend on $\omega_p$ and $\omega_m$,
\begin{equation}\label{13.HT}
M_{l,l'}^{\rm TE}(0)=-\frac{l+1}{l} \tilde{M}_{l,l'}(0),\quad M_{l,l'}^{\rm TM}(0)=\frac{l+1}{l} \tilde{M}_{l,l'}(0).
\end{equation}
The $\Tr\ln(1-M(0))$ in this case is negative at all distances. It is presented in Fig~\ref{Fig3}, right panel, by red thick dash-dotted curve.

At large distances,
\begin{eqnarray}
\label{14.HT}
\Tr \ln(1-\mathbf{M}(0))&\approx & -\frac{1}{2}\Tr \,\mathbf{M^2}(0) \approx -\frac{3}{32}\rho^6+\mathcal{O}\left(\rho^8\right).
\end{eqnarray}
 
In flat geometry, the high temperature limit of the free energy is defined by the zeroth Matsubara  frequency  in the Lifshitz formula~\cite{BKMM}. For the perfectly conducting plate the reflection coefficients entering the Lifshitz formula are 
$r^{cond}_{TM}=1$ and $r^{cond}_{TE}=-1$, while
for the  plate with $\varepsilon$ and $\mu$ both defined by plasma model, in the zero frequency limit the reflection 
coefficients are simplified to  $r^{plasma}_{TM}=1$ and $r^{plasma}_{TE}=1$. Therefore, they do not depend on the parameters 
of the plasma model. The high temperature limit of the free energy density is $f_{||} =-\zeta(3)T/(64\pi d^2)$. At high temperature $f_{||}$ is negative corresponding to attraction.
The free energy in sphere-plate geometry at short distances, $1<<dT<<RT$ and $d/R<<1$, is dominated by the PFA result 
$$\F^{PFA}=2\pi R d f_{||}=-\frac{T}{32}\frac{R}{d}\zeta(3).$$

\subsection{Low temperature}
Here we evaluate the low temperature corrections to the Casimir energy, when the magnetic permeability  $\mu\ne1$ and discuss their influence on the repulsion. 

First we replace the  sum over the Matsubara frequencies~(\ref{1.F2}) by integrals according to
the Abel-Plana formula. The free energy splits into two pieces,
\begin{equation}\label{1.FT} {\cal F}=E_0+\Delta_T {\cal F},
\end{equation}
where $E_0$ is the vacuum energy~(\ref{E1}), and  the temperature dependent part of the free energy is defined by
\begin{equation}\label{1.F2}  \Delta_T {\cal F}  =
\frac{1}{2}\int_{-\infty}^\infty\frac{d\xi}{2\pi}\ n_T(\xi)
\, i \, {\rm Tr}  \left[\ln(1-\mathbf{M}(i\xi))-\ln(1-\mathbf{M}(-i\xi))\right].
\end{equation}

Due to the Boltzmann factor  $n_T(\xi)=1/(\exp(\xi/T)-1)$ in (\ref{1.F2}), for low temperature we may replace the integrand by its  expansion   at $\xi\to0$. For the electromagnetic field, ${M}_{ll',m}$ is a matrix in electromagnetic
polarizations.
Expanding the matrix elements in powers of $\xi$ we obtain
\begin{eqnarray}
\mathbf{M}\equiv&&
\left(\begin{array}{ll} 
\mathbb{M}^{TE}\; \mathbb{M}^{12}\\ \mathbb{M}^{21}
\;\mathbb{M}^{TM} 
\end{array}\right)=\sum_{i=0}^{\infty}{\mathbf{M}}_i \xi^i, \nonumber\\
&=&\left\{\left(\begin{array}{ll}\mathbb{M}^{TE}_0\; 0\\ 0 \quad \mathbb{M}^{TM}_0 \end{array}\right)
+\xi \left(\begin{array}{ll}0\;\; \mathbb{M}^{12}_{1} \\ \mathbb{M}^{21}_{1} \;0 \end{array}\right) 
+
\xi^2\left(\begin{array}{ll}\mathbb{M}^{TE}_2\; 0\\ 0 \;\mathbb{M}^{TM}_2 \end{array}\right)
+\xi^3\left(\begin{array}{ll}\mathbb{M}^{TE}_3\; \mathbb{M}^{12}_3\\ \mathbb{M}^{21}_3\;\mathbb{M}^{TM}_3
\end{array}\right)\right\}+\dots \ {.}
\label{M_exp}
\end{eqnarray}
Here $\mathbb{M}_i$  are matrices over $l,l'$ and diagonal matrices with respect to $m$. 

%
%
The coefficients $\mathbf{M}_i=\mathbf{M}_i(\rho)$ are dimensionless functions of the ratio $\rho=R/L$.
Inserting the expansion  (\ref{M_exp}) into the trace of the logarithm and keeping only the first two odd orders we get
\begin{equation}\label{2.M2}
{\rm Tr} \ln\left(1-\mathbf{M}(\xi)\right)=N_1(\rho)L\xi+N_3(\rho)(L\xi)^3+\dots
\end{equation}
with
\begin{eqnarray}\label{2.N}
N_1&=&-{\rm Tr}\left[ \left(1-\mathbf{M}_0\right)^{-1}\mathbf{M}_1\right], \nonumber \\
N_3&=&-{\rm Tr} \left[\left(1-\mathbf{M}_0\right)^{-1}\mathbf{M}_3\right]
-{\rm Tr} \left[\left(1-\mathbf{M}_0\right)^{-1}\mathbf{M}_1\left(1-\mathbf{M}_0\right)^{-1}\mathbf{M}_2\right]
-\frac{1}{3}\,{\rm Tr} \left[\left(\left(1-\mathbf{M}_0\right)^{-1}\mathbf{M}_1\right)^3\right]\,,
\end{eqnarray}
which are functions of $\rho$ like the $\mathbf{M}_i$'s.

The structure of the expansion directly implies that ${\mathbf{M}}_1(1-{\mathbf{M}}_0)^{-1}=0$.
Consequently,
\begin{eqnarray}
i[\ln(1-{\mathbf{M}}(i\xi))-\ln(1-{\mathbf{M}}(-i\xi))]=-2{\mathbf{M}}_3(1-{\mathbf{M}}_0)^{-1}\xi^3+
{\mathcal O}(\xi^5).
\end{eqnarray}
After taking the trace over polarizations we arrive at
\begin{eqnarray}
N_1^{EM}=0, \quad N_3^{EM}= \sum_{l=1}^{\infty}\sum_{m=-l}^{l}
[{\mathbf{M}}_3^{TE}(1-{\mathbf{M}}_0^{TE})^{-1}
+{\mathbf{M}}_3^{TM}(1-{\mathbf{M}}_0^{TM})^{-1}].
\label{3.N1N3_EM}
\end{eqnarray}
The TE and  TM polarizations decouple in $\xi^3$-approximation. 

Finally the low temperature correction to the Casimir energy in sphere-plate geometry is 
\be
\label{lowT_corr}
\Delta_T \F=\frac{\pi^2 T^4 L^3}{15} N_3(\rho).
\ee

\begin{figure}[b,t]
\epsfig{file= 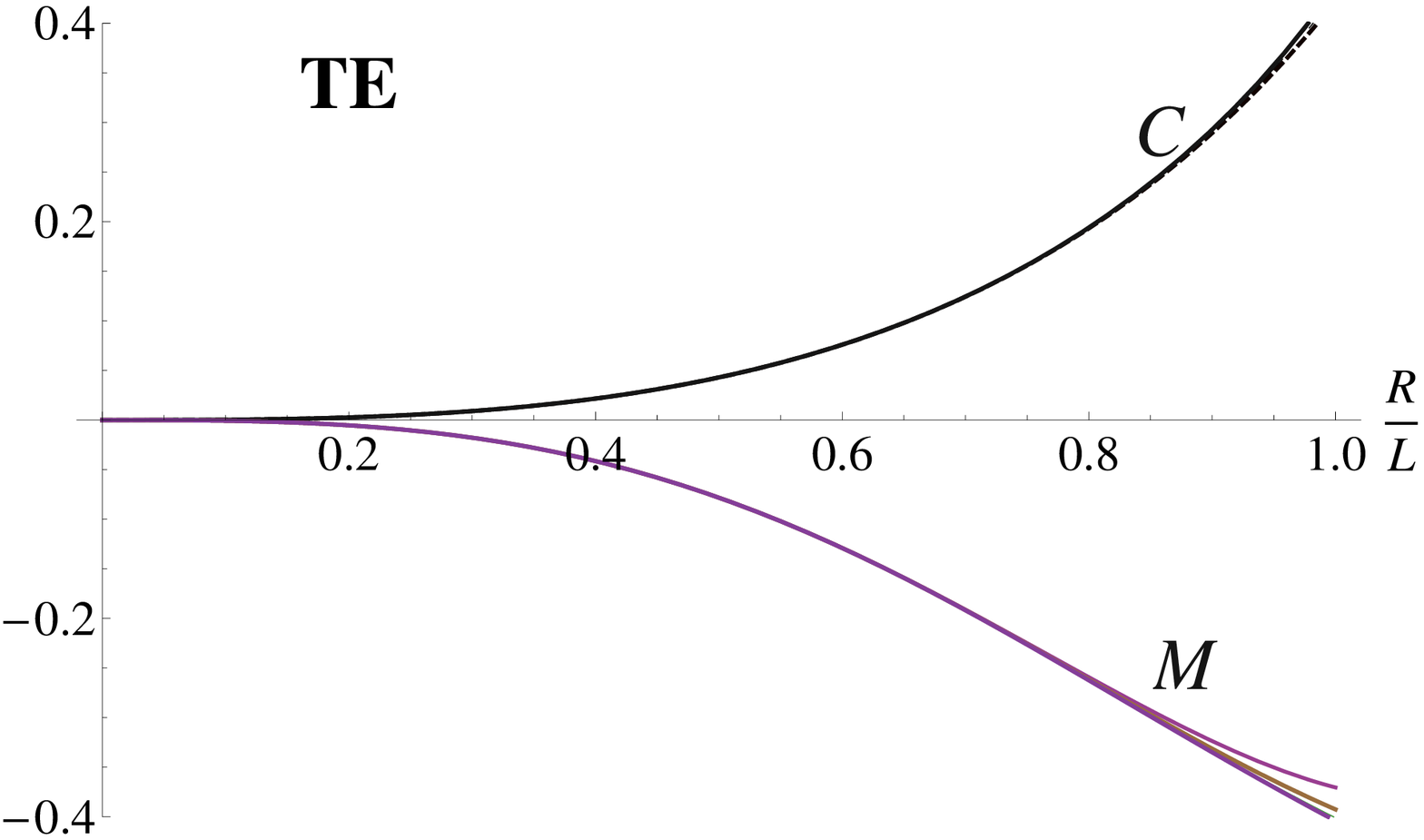, width=7cm}
\epsfig{file= 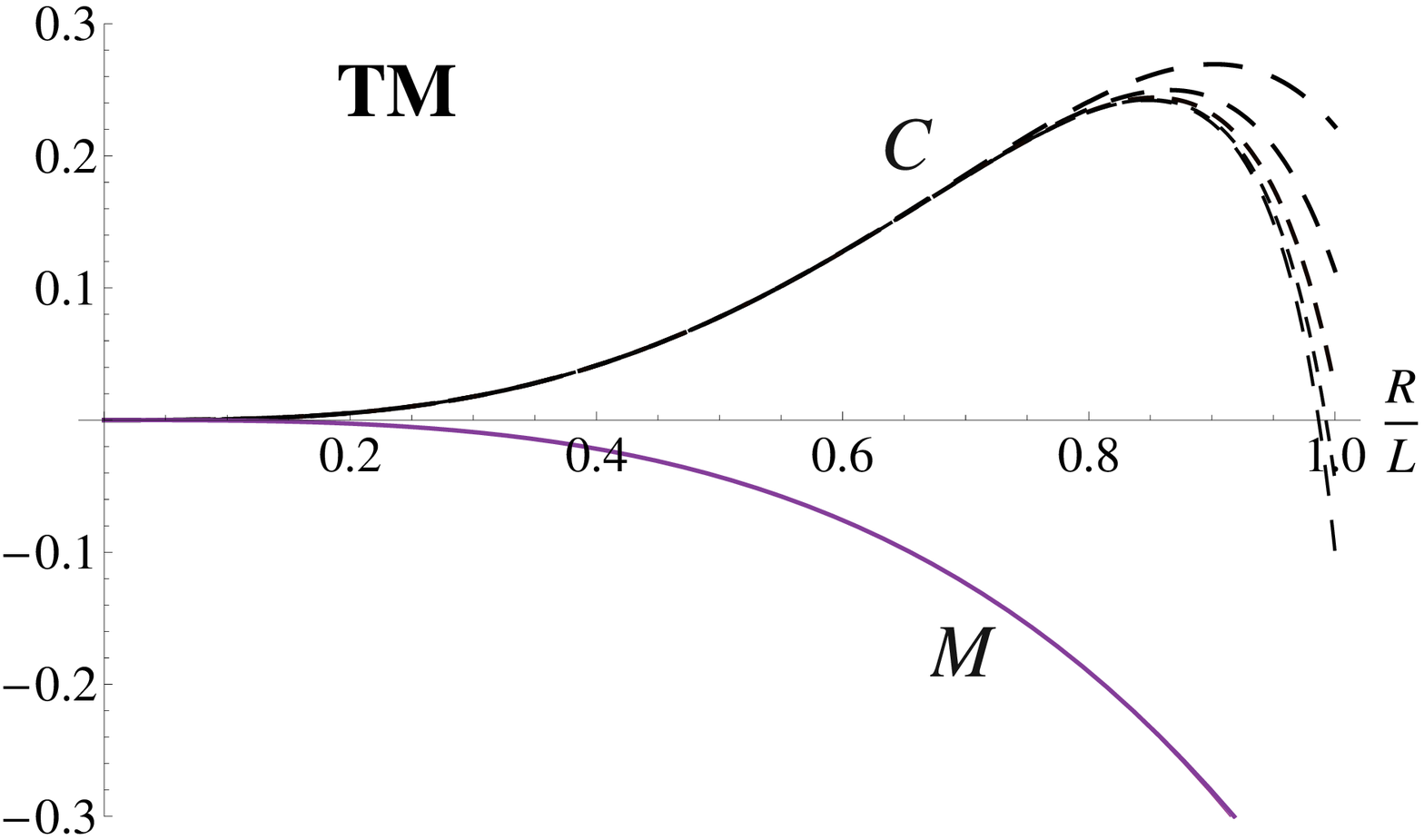,width=7cm}
\caption{The  $N_3^{TE}(\rho)$ and  $N_3^{TM}(\rho)$ for perfectly conducting ball (C) and perfectly permeable ball (M) facing perfectly conducting plate.  The left panel gives the plots for TE modes, the right panel presents the results the  TM modes. $l_{max}$
varies from $1$ to $5$.}
\label{Fig4}
\end{figure}

The  functions $N_3^{TE}(\rho)$ and  $N_3^{TM}(\rho)$ for perfect conductor boundary conditions were obtained numerically in~\cite{Bordag:2010qq}.  The comparison of these results with those obtained for perfectly permeable ball is presented at Fig~\ref{Fig4}. The left panel gives the plots for TE modes, the right panel gives the  TM modes, $R/L$ varies from $0$ (large separation) to $1$ (short separation). To estimate the functions $N_3^{TE}(\rho)$ and  $N_3^{TM}(\rho)$ numerically, one has to truncate the sum over $l$ in (\ref{3.N1N3_EM}) at some $l_{max}$ sufficiently large to stabilize the curves in the pictures. Fig~\ref{Fig4} demonstrates the results with $l_{max}$ varying from $1$ to $5$. For the TE mode, the result weekly depends on $l_{max}$. For the TM mode, as $\rho\to 1$, the result considerably changes with $l_{max}$ and does not stabilize. We discussed this behavior for perfectly conducting ball in~\cite{Bordag:2010qq}. For perfectly permeable ball we observe inverse picture. The results for the TM mode weekly depend on $l_{max}$, while the $TE$ mode  at short distances varies with increasing $l_{max}$.  

\begin{figure}[b]
\epsfig{file=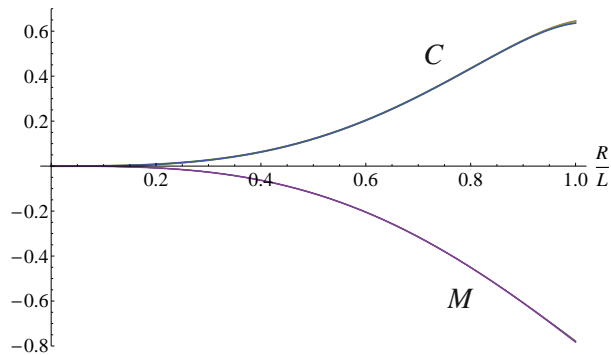,width=8cm}
\caption{The functions $N_3$ for perfectly conducting and perfectly permeable balls. $l_{max}$ changes from $1$ to $10$.}
\label{Fig5}
\end{figure}

Summing up the contributions from $TE$ and $TM$ polarizations at large separations, we derive 
the following  expansion,
\be
\label{N3cond}
N_3^{cond}=\rho^3-\frac{1}{4}\rho^6-\frac{5}{64}\rho^9+\mathcal{O}(\rho^{11}).
\ee
In a similar fashion we obtain the low temperature correction for perfectly  permeable ball above the plane,
 \be
\label{N3magn}
N_3^{magn}=-\rho^3+\frac{1}{4}\rho^6-\frac{1}{32}\rho^9+\mathcal{O}(\rho^{11}).
\ee
The total function $N_3$ comprising  the  sum of TE and TM modes is given on Fig.\ref{Fig5}. Thanks to some cancellations between the TE and TM modes, the total value of $N_3$ is less sensitive to $l_{max}$ than $N_3^{TE}(\rho)$ and  $N_3^{TM}(\rho)$. 
The low-temperature correction to the free energy in the presence of perfectly conducting ball facing perfectly conducting plane
is positive, therefore reducing the attractive force between the sphere and the plate. When the ball is perfectly permeable,
the correction is negative, making the repulsion weaker.  

In the present paper we shall not calculate low temperature corrections to the Casimir energy in all the cases~(\ref{5.HT}).
For the sphere with with constant dielectric permittivity and magnetic permeability, case ({\bf a}), the corrections were studied
in~\cite{Bordag:2010qq}. It was shown that $N_3$ is  positive when $\varepsilon>\mu$, and $N_3$ is negative when $\varepsilon<\mu$. 
Thus at low temperature both the attraction and the repulsion are getting weaker than at zero temperature.

\section{Discussion}
In the present paper we set the limits on the repulsion in sphere-plate geometry.  The strongest possible repulsion is achieved between perfectly conducting plate and perfectly permeable sphere. When these approach each other, the corresponding Casimir energy tends to  $-7/8$ of the PFA result for perfectly conducting sphere facing perfectly conducting plate. In the case of repulsion, we have found the first analytic  correction having the same sign as the leading term. At large separations, $\rho\to0$, the  energy is given by an asymptotic series in powers of  $\rho$ starting from $\rho^3$. We conclude that the ratio of the strongest repulsive vacuum energy to strongest attractive depends on the separation. It equals to $-1$ when the bodies are far apart, and tends to -7/8 in close proximity.  

We have studied the influence of  finite conductivity and finite temperature on the repulsion in the given geometry.  Regardless of the model used for the dielectric permittivity or magnetic permeability, the vacuum energy never exceeds the limiting values set in sphere-plane geometry with boundary conditions of perfect conductor or perfect magnetic on the sphere.
We have derived the leading low temperature corrections to these limiting values. The corrections are proportional to $T^4$ and reduce both, attraction and repulsion. The diminishing of the attraction by the thermal photons was previously observed in~\cite{PhysRevA.82.012511}.

At high temperature the free energy of the electromagnetic field in the presence of perfectly conducting plane and perfectly permeable sphere is positive resulting in repulsion. For a sphere with constant dielectric permittivity and magnetic permeability 
the free energy is negative when $\varepsilon\geq\mu$. It is positive  when  $\varepsilon<<\mu$. If $\varepsilon$ is slightly lower than $\mu$ the sign of the fee energy depends on the distance, changing from positive at large distances to negative at moderate separations. 

For both, dielectric permittivity and magnetic permeability defined by plasma model, the high temperature limit of the free energy is negative ensuring attraction at all separations. Similarly the free energy is negative  when the dielectric permittivity is defined by plasma model and constant magnetic permeability.
The free energy may vary from positive at large and medium distances to negative at close separations when the magnetic permeability 
is defined by plasma model and constant dielectric permittivity. For a given magnetic plasma frequency $\Omega_m$ the free energy decreases with increasing $\varepsilon$. Summarizing, at high temperature the ratio between the strongest possible repulsion and the strongest possible attraction depends on the distance between the sphere and the plate, varying from -3/4 at short separations to -1 at long separations.

The transition from an attractive to a repulsive regime for parallel plates with  $\varepsilon_i,\mu_i=const$, $i=1,2$, was studied in~\cite{PhysRevLett.89.033001}. It was shown, that in the high temperature limit the force is attractive for any values of permittivity and permeability, $\varepsilon_i,\mu_i>1$. Hence the force between parallel plates, which at zero temperature is  repulsive for given values of $\varepsilon_i$ and $\mu_i$, $i=1,2$,  changes its sign as the system is heated.  
In contrast to this, in  sphere-plane geometry we observe, that the force between a perfectly conducting plane and a ball remains repulsive for both zero and high temperatures, Fig.~\ref{Fig2}, for certain values of  $\mu$ and $\varepsilon$ corresponding to the ball.

\section*{Acknowledgments}
The work was supported by Russian Foundation of Basic Research (Grant N 11-02-12232-ofi-m-2011) and the Heisenberg-Landau Program (Project HLP-2012-33).
 

\end{document}